\documentclass{svmult}
\begin{document}
%%%%%%%%%%%%%%
\title*{Action Ward Identity and the St\"uckelberg-Petermann renormalization 
group\thanks{Talk given at the Symposium in honour of the 70th anniversary 
of Jacques Bros, 
Paris 2004, based on \cite{DF3} and on private communication with {R.~Stora}}}
%%%%%%%%%%%%%%
\author{Michael D\"utsch\inst{1} \and Klaus Fredenhagen\inst{2}}
 \institute{Institut f\"ur Theoretische Physik,
Universit\"at Z\"urich,
CH-8057 Z\"urich, Switzerland,
{\tt \small duetsch@physik.unizh.ch}\\
 \and 
II. Institut f\"ur Theoretische Physik,
Universit\"at Hamburg,
D-22761 Hamburg, Germany,
{\tt \small klaus.fredenhagen@desy.de}}
\titlerunning{Action Ward Identity and renormalization group}
\maketitle
\begin{abstract}
A fresh look at the renormalization group 
(in the sense of St\"uckelberg-Petermann) from the point of view of 
algebraic quantum field theory is given, and it is shown that a consistent 
definition of local algebras of observables and of interacting fields 
in renormalized perturbative quantum field theory can be given in terms 
of retarded products. The dependence on the Lagrangian enters this
construction only through 
the classical action. This amounts to the commutativity of retarded 
products with derivatives, a property named Action Ward Identity by Stora. 
\end{abstract}
\section{Introduction}
Modern perturbative quantum field theory is mainly based on path integrals. 
Basic objects are the correlation functions
%%%%%%%%%%%%%
\begin{equation}
      G(x_{1},\ldots,x_{n})= 
        (\Omega,T\varphi(x_{1})\cdots\varphi(x_{n}) \Omega) \ 
\end{equation}
%%%%%%%%%%%%%
which are calculated as moments of the Feynman path integral, i.e.
%%%%%%%%%%%%%
\begin{equation}
      G(x_{1},\ldots,x_{n}) = 
       \frac{1}{Z}\int{\cal D}\varphi \, \varphi(x_{1})\cdots 
        \varphi(x_{n})e^{iS(\varphi)/\hbar} \ .
\end{equation}
%%%%%%%%%%%%
Relations to Wightman fields or even to local algebras of 
observables are indirect and are not elaborated in typical cases.
This fact leads to some severe problems. On the one hand side, concepts 
from the algebraic approach to quantum field theory cannot be 
easily applied to 
perturbatively constructed models. This lead to the prejudice that these 
concepts are irrelevant for physically interesting models as long as 
these can be constructed only at the level of formal perturbation theory.
On the other side, the path integral approach is intrinsically nonlocal; 
therefore difficulties arise for the treatment of infrared problems and of 
finite temperatures, and theories on curved back grounds or with 
external fields cannot properly be formulated. In algebraic quantum field 
theory, concepts for dealing with these problems have been developed
\cite{Haag,BB,BF,HW1-2,BFV,Mar}.

We therefore started a program for a perturbative construction of the 
net of local algebras of observables \cite{BF,DF1}. There already exists a local formulation
of renormalization due to Epstein and Glaser \cite{EG} and based on older ideas of 
St\"uckelberg and Bogoliubov \cite{BS}.
The advantages of this method compared to other schemes of renormalization 
are that it can be formulated entirely in position space, that it is 
mathematically well elaborated and that it gives a direct construction of 
operators. It is, however, not equally well developed from the computational 
point of view, the r\^{o}le of the renormalization group is not yet fully 
established and, even worse, the application to non-Abelian gauge theories 
is not evident.

In particular the last problem was partially resolved by the Z\"urich school
\cite{Scharf,S-wiley}, who also developed many new tools for doing concrete 
calculations in the Epstein-Glaser framework.

The starting point for our approach is Bogoliubov's definition of interacting 
fields
%%%%%%%%%%%%%%
\begin{equation}
       \varphi_{\int\!{\cal L}g}(x) = 
        (Te^{i\!\int\!{\cal 
        L}g})^{-1}\,\frac{\delta}{\delta h(x)} 
        Te^{i(\!\int\!{\cal L}g+\int\!\varphi h)}|_{h=0}
\end{equation}
In this contribution, we will first formulate axioms for interacting fields,
will then construct solutions, thereafter discuss the renormalization group 
and will finally describe the arising local nets and local fields.
%%%%%%%%%%%%%%%%%%%%%%%%%%%%%%%%%%%%%%%%%%
\section{Basic properties required for interacting fields}
%%%%%%%%%%%%%%%%%%%%%%%%%%%%%%%%%%%%%%%%%%%
We consider polynomial functionals of a classical scalar 
  (${\mathcal C}^{\infty}$) 
field $\varphi$ on $d$-dimensional Minkowski space,
%%%%%%%%%%%%
\begin{equation}
F(\varphi)=\sum_{n} \langle f_{n},\varphi^{\otimes 
        n}\rangle \ . 
\end{equation}
%%%%%%%%%%%%
Here the smearing functions
$f_{n}$ are assumed to be distributions with compact support in $n$ 
variables 
with wave front set
%%%%%%%%%%%%
\begin{equation}
        WF(f_{n})\subset \{(x,k)\>,\>\sum_i k_{i}=0\}
\end{equation}
%%%%%%%%%%%%
e.~g. $f_{2}(x_{1},x_{2})=g(x_{1})\delta(x_{1}-x_{2})$ 
with $g\in\mathcal{D}$, so 
$\langle f_2,\varphi^{\otimes 2}\rangle =\int dx\, g(x)\varphi(x)^2$.

A functional $F$ is called local, if 
$\mathrm{supp}(f_{n})\subset D_{n}$, with the total diagonal
\begin{equation}
D_{n}=\{(x_{1},\ldots,x_{n})\>,\>x_{1}=\ldots =x_{n}\in{\bf R}^d\} \ .\label{diagonal}
\end{equation}
%%%%%%%%%%%%%%%%%%%%%%%%%%%%%%%%%%%%%
Let us first describe the classical free theory: There the
field equation $(\qed +m^2) \varphi=0$ generates an ideal 
in the algebra of functionals $F$ (with respect to 
pointwise multiplication).
The Poisson bracket of the classical model is
  %%%%%%%%%%%%%
  \begin{equation}
       \{F,G\}(\varphi)=\int dxdy\frac{\delta F}{\delta\varphi(x)} 
        \Delta(x-y) \frac{\delta G}{\delta\varphi(y)}
  \end{equation}
  %%%%%%%%%%%%
where $\Delta$ is the commutator function of the free scalar 
field with mass $m$.
  
For quantization we follow the recipes of deformation quantization
and define an associative product $*$ by
  %%%%%%%%%%%
  \begin{equation}
      F*G(\varphi) = 
        e^{\hbar\!\int\! dx dy 
        \Delta_{+}(x-y)\frac{\delta}{\delta\varphi(x)} 
        \frac{\delta}{\delta\varphi'(y)}} F(\varphi) 
        G(\varphi')|_{\varphi'=\varphi}
   \end{equation}
   %%%%%%%%%%%%%
where $\Delta_{+}$ is the 2-point function of the free scalar field.
   
The Poisson bracket as well as the $*$-product vanish on the ideal 
generated by the field equation and can therefore be defined also on the 
quotient algebra. In the classical case one obtains the Poisson algebra of 
the classical field theory, in the quantum case one obtains the algebra 
of Wick polynomials on Fock space, and the formula for the $*$-product 
translates 
into Wick's Theorem. For the treatment of interactions, it is however 
preferable not to go to the quotient but to formulate everything on the 
original space (off shell formalism). This introduces some redundancy which 
will be very useful for the solution of cohomological problems, as for 
instance in the determination of counter terms in the Lagrangian which 
compensate changes in the renormalization prescription.  

We now turn to the characterization of retarded interacting field functionals.
Let $F,S_n$, $n\in {\bf N}$ be local functionals of $\varphi$.
Let $S(\lambda)=\sum_{n=1}^{\infty}\lambda^n S_n$ be a formal power series 
with vanishing term of zeroth order. 
We associate to the pair $F,S(\lambda)$ a formal power series of 
functionals $F_{S}$ which we interpret as the functional $F$ of the 
interacting retarded field under the influence of 
the interaction $S$ where $\lambda$ is the expansion parameter 
of the formal power series.
   
We require the following properties:
  \begin{description} 
        \item[Initial condition:]  
      \begin{equation}
      F_{S=0}=F
      \end{equation}
        \item[Causality:] 
        \begin{equation}F_{S_{1}+S_{2}}=F_{S_{1}}\end{equation} 
    if $S_{2}$ takes place later than $F$.
   
        \item[Glaser-Lehmann-Zimmermann:] \cite{GLZ}  
    \begin{equation}\frac{i}{\hbar}[F_{S},G_{S}]_\star = 
         \frac{d}{d\mu}(F_{S+\mu G}- G_{S+\mu F})|_{\mu=0 } \ . \end{equation}  
   \end{description} 
In addition, we require
\begin{description}
\item[Unitarity:] \begin{equation}(F_S)^*=F^*_{S^*}\end{equation}

\item[Covariance:] Let $\beta$ denote the natural action of the 
      Poincar\'{e} group on the space of functionals $F$. Then
      \begin{equation}\beta_L(F_S)=\beta_L(F)_{\beta_L(S)}\end{equation}
      for all Poincar\'{e} transformations $L$.
\item[Field independence:] The association of local functionals to 
      their interacting counterpart should not explicitly depend on 
      $\varphi$,
      \begin{equation}
      F_S(\varphi +\psi)= F(\varphi+\psi)_{S(\varphi +\psi)}
      \end{equation}
      for test functions $\psi$.

\item[Scaling:] There is a natural scaling transformation, $x\mapsto\rho x$,
    on the 
    space of functionals which scales also the mass in the 
    $*$-product from 
    $m$ to $\rho^{-1} m$. But the limit $\rho\to \infty$ 
    is singular because of 
    scaling anomalies. This holds already for the free theory 
    (in even dimensions).
    
    Instead: Introduce an auxiliary mass parameter $\mu>0$ 
    and transform the 
    $*$-product to an equivalent one which is smooth in the mass
    $m$.
    Require a smooth $m$ dependence in the prescription 
    $(F,S)\to F_S^{(m,\mu)}$ (in particular at $m=0$ 
    (one-sided)) such that 
    $F_S^{(\rho^{-1}m,\mu)}$, scaled by $\rho$, is in 
    every order of perturbation theory a 
    polynomial of $\log \rho$.
  \end{description}
\section{Construction of solutions}
We make the
ansatz that the interacting field is a formal power series in the interaction, 
\begin{equation}F_{\lambda S}=\sum_{n=0}^{\infty}\frac{\lambda^n}{n!} 
R_{n,1}(S^{\otimes n}\otimes F) \ ,\end{equation}
where $R_{n,1}$ is an ($n+1$)-linear functional, 
symmetric in the first 
$n$ entries (called 'retarded product').

The inductive construction of retarded products can be done in 
an analogous way as the construction of time ordered products in the 
Epstein-Glaser method and was elaborated already 
(in a slightly different form)
by Steinmann \cite{Ste}.
In a first step we represent the local functionals as fields smeared 
with a test function,
\begin{equation}F(\varphi)=\int d^dx\, A(x)f(x)\end{equation}
where $A$ is an element of the space ${\cal P}$ of all 
polynomials of $\varphi$ and its derivatives,
\begin{equation}A(x)=A(\varphi(x),\partial\varphi(x),\ldots) \ ,\end{equation}
and $f$ is a test function. 
Using this representation,
the retarded products can be represented as distributions 
$R_{n,1}(A_1(x_1),\ldots,A_{n+1}(x_{n+1}))$
with values in 
the space of functionals of $\varphi$.
With that the inductive construction can be done by first constructing 
the retarded products outside of the total diagonal $D_{n+1}$ (\ref{diagonal})
in terms of retarded products with less arguments and 
then by extending them to the full space (in the sense of 
distributions). Perturbative renormalization is precisely this extension problem.

A difficulty with this procedure is that the representation of local 
functionals by fields is non-unique, e.g. one finds by 
partial integration
\begin{equation}
\int dx\, \partial A(x)f(x)= -\int dx\, A(x) \partial f(x)\ .
\end{equation}

Therefore it was suggested by Stora \cite{AWI} that one should impose an 
additional requirement, termed the {\bf Action Ward Identity} (AWI), which guarantees 
that the retarded products depend only on the functionals, but not on 
the way they are represented as smeared fields. This amounts to the 
requirement, that the retarded products commute with derivatives, 
\begin{equation}\partial 
R_{n,1}(\ldots,A(x),\ldots)=R_{n,1}(\ldots,\partial 
A(x),\ldots) \ .\end{equation}
Such a relation cannot hold  
if the arguments of the retarded products are on-shell fields
(i.e.~satisfy the free field equations).
In the off-shell formalism adopted here, however,        
one can actually show that there are no 
anomalies for the AWI. Namely one may 
introduce {\bf balanced} fields, as was done recently for the purpose of 
non-equilibrium quantum field theory by Buchholz, Ojima and Roos \cite{BOR}:
the balanced fields form that subspace ${\cal P}_{\rm bal}$ of all local fields
\begin{equation}A(x)=
P(\partial_1,\ldots,\partial_n)
\varphi(x_1)\cdots\varphi(x_n)|_{x_1=\ldots=x_n=x}\in {\cal P}\end{equation}
(where $P$ is a polynomial) which arises when $P$ is restricted
to depend only on the differences of variables $(\partial_i-\partial_j)$. 

The argument relies on
\begin{lemma} 
Every local functional $F$ is of the form
             \begin{equation} F=\int d^dx\, f(x)\end{equation}
with a unique test function $f$ with values in the space of 
balanced fields.
\end{lemma}

\begin{proof} {\it Existence:} Every symmetric polynomial $P(p_1,\ldots,p_n)$ in 
$n$ variables $p_1,\ldots,p_n$,
 $p_i\in {\bf R}^d$ may be written in the form
\begin{equation}P(p_1,\ldots,p_n)=\sum_j a_j(p)b_j(p_{\mathrm{rel}})\end{equation}
with polynomials $a_j$ in 
the center of mass momentum $p=\sum p_i$ and symmetric 
polynomials $b_j$ in the 
relative momentum $p_{\mathrm{rel}}=(p_i-p_j,i<j)$. 
Every local functional of $n$-th order in 
$\varphi$ is of the form
\begin{equation}
F(\varphi)=\int d^dx \sum_{j}\lambda_{j}(x)
\end{equation}
\begin{equation}
\times\int d^{d(n-1)}x_{\mathrm{rel}}\,\delta(x_{\mathrm{rel}})
a_j(i\partial)\,
b_j(i\partial_{\mathrm rel})
\varphi^{\otimes n}(x,x_{\mathrm{rel}})
\end{equation}
with test functions $\lambda_{j}$. By partial integration we find
\begin{equation}
F(\varphi)=\int dx\, \sum_j \mu_j(x) B_j(x)
\end{equation}
with $\mu_j=a_j(-i\partial)\lambda_{j}$ and 
$B_j=b_j(i\partial_{\mathrm{rel}})\varphi^{\otimes n}|_{x_{\mathrm{rel}}=0}
\in {\cal P}_{\rm bal}$.      

\noindent {\it Uniqueness:} To show uniqueness let
\begin{equation}
F(\varphi)=\int dx \sum_j \mu_j(x) B_j(x)
\end{equation}
with the balanced fields
$B_j=
b_j(i\partial_{\mathrm{rel}})\varphi^{\otimes n}|_{x_{\mathrm{rel}}=0}
$, where $b_j$ are {\it symmetrical} polynomials (with respect to permutations 
of $\partial_1,...,\partial_n$). Then $F(\varphi)=0$ implies that 
\begin{equation}
\sum_j\mu_j(x) b_j(-i\partial_{\mathrm{rel}})
\delta(x_{\mathrm{rel}})=0 \ .
\end{equation}
Since the center of mass coordinate $x$ and the relative coordinate 
$x_{\mathrm{rel}}$ are independent, 
this implies
\begin{equation}
\sum_j\mu_j\otimes b_j=0 \ ,
\end{equation}
hence $\sum_j\mu_j B_j =0$. \qed
\end{proof}
By using this Lemma the AWI can be fulfilled by the following procedure. One first 
extends $R(A_1(x_1),...,A_{n+1}(x_{n+1}))$ only for $A_1,...,A_{n+1}\in {\cal P}_{\rm bal}$.
Since, by induction, the AWI holds outside of $D_{n+1}$, one may then define 
the extension for general fields $A_1,...,A_{n+1}$ by using the AWI and 
linearity in the fields.
%%%%%%%%%%%%%%%%%%%%%%%%%%%%%%%%%%%%%%%%
\section{Renormalization group}
%%%%%%%%%%%%%%%%%%%%%%%%%%%%%%%%%%%%%%%%
The extension of $R_{n,1}$ from ${\cal D}({\bf R}^{d(n+1)}\setminus D_{n+1})$
to ${\cal D}({\bf R}^{d(n+1)})$ is generically non-unique. Hence, there is an
ambiguity in the construction of interacting fields, which is well understood:
it can be described in terms of the St\"uckelberg-Petermann 
   renormalization group ${\cal R}$ \cite{StPe}. 
   
   The elements of ${\cal R}$ are analytic invertible maps 
   $Z:S\mapsto Z(S)$ which map the space of formal power series of local 
   functionals, which start with the first term, into itself such that
\begin{eqnarray}
  & Z(0)=0\ , &\\
  & Z'(0)={\rm id} &\quad (\mathrm{where}\>\>
   Z^\prime (S)F:=\frac{d}{d\tau}Z(S+\tau F)\vert_{\tau=0})\ ,\\
  & Z(S^\star)=Z(S)^\star &
\end{eqnarray}
and
   \begin{equation}Z(S)=\int d^dx\, z({\cal L}(x),\partial {\cal L}(x),\ldots)\end{equation} 
   if $S=\int dx\, {\cal L}(x)$.  Here the Lagrangians are of the form 
   ${\cal L}(x)=\sum_j A_j(x)g_j(x)$ 
   with $A_j\in {\cal P}_{\rm bal}$ and $g_j\in {\cal D}({\bf R}^{d})$. Derivatives are defined by 
   $\partial {\cal L}:=\sum_j A_j(x)\partial g_j$. $z$ is of the form
\begin{equation} z({\cal L}(x),\partial {\cal L}(x),\ldots)=\sum_{n,a}\frac{1}{n!}\,
d_{n,a}(A_{j_1}\otimes...\otimes A_{j_n})(x)\,\prod_{i=1}^n\partial^{a_i}g_{j_i}(x)\end{equation}
with linear maps $d_{n,a}:{\cal P}_{\rm bal}^{\otimes n}\rightarrow {\cal P}_{\rm bal}$
which are Lorentz invariant, maintain homogeneous scaling of the fields
and do not explicitly depend on $\varphi$.
   
   The Main Theorem of Renormalization \cite{SP,Pinter,DF3} amounts to the 
   following relation between
   interacting field functionals for two renormalization prescriptions 
   $(F,S)\to F_S$ and $(F,S)\to\hat{F}_S$ which both satisfy the mentioned axioms: 
   \begin{theorem}
   There is a unique element $Z$of the renormalization group  ${\cal R}$ such that
\begin{equation}
\hat{F}_{S}=(Z'(S)F)_{Z(S)} \ .\label{mainthm} 
\end{equation}

Conversely, given a renormalization prescription $F_S$
satisfying the axioms and an arbitrary $Z\in {\cal R}$, equation (\ref{mainthm})
gives a new renormalization prescription $\hat{F}_S$ which fulfills also the axioms.
   \end{theorem}
%%%%%%%%%%%%%%%%%%%%%%%%%%%%%%%%%%%%%%%%%%%%%%
\section{Local nets and local fields}
%%%%%%%%%%%%%%%%%%%%%%%%%%%%%%%%%%%%%%%%%%%%%%
   We first define the algebra  ${\cal A}_{\cal L}(\cal O)$
   of observables within the region ${\cal O}$ for a (fixed) interaction 
   ${\cal L}$ with compact support. 
   This algebra is generated by 
   elements $F_{S}$, $S=\int dx\, {\cal L}(x)$ with 
   local functionals $F$ 
   fulfilling ${\rm supp}\>\frac{\delta F}{\delta\varphi}\subset {\cal O}$.
In \cite{BF} it has been found that the algebraic structure of ${\cal A}_{\cal L}({\cal O})$
is independent of the values of ${\cal L}$ outside of ${\cal O}$:
%%%%%%%%
   \begin{theorem}
     If the interactions ${\cal L}_1$ and ${\cal L}_2$ coincide 
     within ${\cal O}$, there exist 
     isomorphisms $\alpha:{\cal A}_{\mathcal{L}_1}({\cal O})\rightarrow {\cal A}_{\mathcal{L}_2}
     ({\cal O})$ such that
\begin{equation}
     {\alpha}(F_{S_1})=F_{S_2}\label{adlim}
\end{equation}
     for all $F$ localized in ${\cal O}$ (where $S_j=\int dx\, {\cal L}_j(x)$).
   \end{theorem}
%%%%%%%%%%  
   We now want to construct the algebras for a not necessarily compactly 
   supported Lagrangian ${\cal L}$ (i.e.~we perform the 
   so called algebraic adiabatic limit).   
   For this purpose we consider the
   bundle of algebras ${\cal A}_{{\cal L}_1}(\cal O)$ over the space of 
   compactly supported Lagrangians ${\cal L}_1$ which coincide 
   within ${\cal O}$ with the 
   given Lagrangian ${\cal L}$.

   A section $B=(B_{{\cal L}_1})$ of this bundle is called covariantly 
   constant if for all 
   automorphisms $\alpha$ satisfying (\ref{adlim}) the following relation holds:
   \begin{equation}\alpha (B_{{\cal L}_1})=B_{{\cal L}_2} \ .\end{equation}
   So in particular the interacting field functionals $F_S$ are 
   (by definition) covariantly
   constant sections.
   The covariantly constant sections now generate the local algebras 
   associated to 
   the interaction ${\cal L}$. 

   To define the net of local algebras one has to specify, for 
   ${\cal O}_1\subset {\cal O}_2$, 
   the injections
    \begin{equation}i_{{\cal O}_2{\cal O}_1}: 
    {\cal A}_{\cal L}({\cal O}_1) \to {\cal A}_{\cal L}({\cal O}_2) \ .\end{equation}
   Let $B\in{\cal A}_{\cal L}({\cal O}_1)$. Then 
   $i_{{\cal O}_2{\cal O}_1}(B)$ is the section which is obtained from 
   the section $B$ by restriction to Lagrangians which coincide 
   with ${\cal L}$ on the larger region.
   Clearly these 
   injections satisfy the compatibility relation required for nets:
  \begin{equation} i_{{\cal O}_3{\cal O}_2}\circ i_{{\cal O}_2{\cal O}_1}
  =i_{{\cal O}_3{\cal O}_1}\end{equation}
  for ${\cal O}_1\subset {\cal O}_2\subset {\cal O}_3$.
  Moreover, the net satisfies local commutativity and,  if ${\cal L}$ is Lorentz invariant,
  it is covariant under the Poincar\'{e} group \cite{DF1}.

  The next step is the construction of local fields 
  associated to the local net. Following \cite{BFV},
  a local field associated to the net 
  is defined as a family of distributions
  $(A_{\cal O})$ with values in ${\cal A}_{\cal L}(\cal O)$  such that
  \begin{equation}i_{{\cal O}_2{\cal O}_1}(A_{{\cal O}_1}(h))= A_{{\cal O}_2}(h)\end{equation}
  if the test function $h$ has support contained in 
  ${\cal O}_1$.
  In particular all classical fields $A\in {\cal P}$ induce local fields 
  $A^{\cal L}\equiv (A^{\cal L}_{\cal O})_{\cal O}$ by the sections
\begin{equation}
A^{\cal L}_{\cal O}(h)\> :\>
{\cal L}_1\mapsto (A_{\cal O}(h))_{{\cal L}_1}=(\int dx \, A(x)h(x))_{S_1}\label{locfield}
\end{equation}
   where ${\cal L}_1$ coincides with ${\cal L}$ within ${\cal O}$,
   $h\in {\cal D}({\cal O})$ and $S_1=\int dx\,{\cal L}_1$. It is an
   open question whether there are other local fields. 
   The answer would amount to the determination of the Borchers class
   for perturbative quantum field theories.
   
   We may now restrict the
   renormalization group flow to constant Lagrangians 
   ${\cal L}\in {\cal P}_{\rm bal}$.
   %%%%%%%%%%%%%%
    \begin{theorem} Let ${\cal L}\mapsto {\cal A}_{{\cal L}}$ and ${\cal L}\mapsto
      \hat{{\cal A}}_{\cal L}$ (${\cal L}\in {\cal P}_{\rm bal}$) be associations of 
      local nets which are defined by the two renormalization prescriptions
      $F_S$ and $\hat F_S$, respectively.
\begin{enumerate}
        \item Then there exists a unique map ('renormalization 
of the interaction')
\begin{equation}
z_0:\lambda\mathcal{P}_{\mathrm{bal}}[[\lambda]]
\longrightarrow \lambda\mathcal{P}_{\mathrm{bal}}[[\lambda]]\> :
\> z_0(\lambda {\cal L})=\lambda {\cal L}+O(\lambda^2).
\label{z:map}
\end{equation}
such that the nets $\hat{{\cal A}}_{\cal L}$ and ${\cal A}_{z_0({\cal L})}$
are equivalent for all ${\cal L}$.
\item Furthermore there exists a unique map ('field renormalization')
%%%%%%%%%%%%%%%%%%%%
\begin{equation}
z^{(1)}:\lambda\mathcal{P}_{\mathrm{bal}}[[\lambda]]\times
\mathcal{P}[[\lambda]]
\longrightarrow \mathcal{P}[[\lambda]]\>:\>
(\lambda {\cal L},A)\mapsto z^{(1)}(\lambda {\cal L})A
=A+O(\lambda)
\end{equation}
%%%%%%%%%%%%%
such that $z^{(1)}(\lambda {\cal L}):\mathcal{P}[[\lambda]]
\longrightarrow \mathcal{P}[[\lambda]]$ is a linear invertible map which 
commutes with partial derivatives and such that for the local fields (\ref{locfield}) 
the identification of $\hat{{\cal A}}_{\cal L}$ with ${\cal A}_{z_0({\cal L})}$
is given by
%%%%%%%%%%%%%%%%%%
\begin{equation}
\hat{A}^{\cal L}=
(z^{(1)}({\cal L})A)^{z_0({\cal L})}\ ,
\quad\forall {\cal L}\in \mathcal{P}_{\mathrm{bal}},\>\>A\in{\cal P}\ .\label{z^1}
\end{equation}
%%%%%%%%%%%%%%%%%%%%
 \end{enumerate}
\end{theorem}
%%%%%%%%%%%%
We point out that $z_0$ and $z^{(1)}$ are independent of ${\cal O}$.
%%%%%%%%%%%%%
\begin{proof}
By applying the algebraic adiabatic limit to Theorem 2 the formula
(\ref{mainthm}) goes over into (\ref{z^1}) with
%%%%%%%%%%%
\begin{equation}
z_0({\cal L})=z({\cal L},0)=\sum_n\frac{1}{n!}\,d_{n,0}({\cal L}^{\otimes n})
\end{equation}
%%%%%%%%%%%%%%%%%%%%
and
%%%%%%%%%%%%%%%%%%
\begin{equation}
z^{(1)}({\cal L})A=\sum_{n,a\in {\bf N}_{0}^{d}}\frac{1}{n!} 
\,(-1)^{|a|}\,\partial^a d_{n+1,(0,\ldots,0,a)}
({\cal L}^{\otimes n}\otimes A)\ .\label{z^1explicit}
\end{equation}
%%%%%%%%%%%%%%%%%%%%
For the equivalence of the nets we refer to Theorem 5.1 of \cite{DF3}. \qed
 \end{proof} 
%%%%%%%%%%%

\noindent {\it Example:} ${\cal L}=\varphi^4$ in $d=4$ dimensions. Then 
\begin{equation}
z_0{\cal L}=(1+a) \varphi^4 + 
   b ((\partial \varphi)^2-\varphi\qed\varphi) + m^2 c \varphi^2 +m^4 e\ ,
\end{equation}
where $a,b,c,e\in {\bf C}[[\lambda]]$. In the case $m=0$ the coupling constant 
renormalization $a$ and the field renormalization $b$ are explicitly computed to 
lowest non-trivial order in \cite{DF3}.
\medskip
        
\noindent  {\it Remarks.} 
(1) It may happen that for a certain interaction ${\cal L}_0$ the renormalization 
is trivial, $z_0({\cal L}_0)={\cal L}_0$, but the corresponding field renormalization
$z^{(1)}({\cal L}_0)$ is non-trivial. This corresponds to the Zimmermann relations.

   (2) The scaling transformations on a given renormalization 
   prescription induce a one parameter subgroup of the renormalization 
   group, which may be called a Gell-Mann-Low Renormalization Group. 
   Its generator is related to the $\beta$-function. 
   Gell-Mann-Low subgroups belonging to different 
   renormalization prescriptions are 
   conjugate to each other. The generator starts with a term of 
   second order which is universal \cite{BDF}.   

   (3) An analysis of the perturbative renormalization group in the algebraic 
   adiabatic limit was already given by Hollands and Wald in the more general 
   framework of QFT on curved space times \cite{HW}. 
   But the formalism presented here is not 
   yet fully adapted to general Lorentzian space-times.

   (4) Hollands and Wald generalized the Action Ward Identity to curved 
    space times \cite{HW4} (and called it 'Leibniz rule'). In that 
    framework it is a 
    non-trivial condition already for time ordered (or retarded) 
    products of {\it one} factor.  

   (5) Ward identities, in particular the Master Ward Identity \cite{BD,DF2} 
   remain to be analyzed.


\begin{thebibliography}{999}

\bibitem{BS}Bogoliubov, N.N., and Shirkov, D.V., {\it "Introduction to the 
Theory of Quantized Fields"}, New York (1959)

\bibitem{BB} Bros, J., and Buchholz, D., ``Towards a Relativistic KMS Condition'',
{\it Nucl.~Phys.} {\bf B429} (1994) 291-318

\bibitem{BDF}Brunetti, R., D\"utsch, M., and Fredenhagen, K., work in progress

\bibitem{BF}Brunetti, R., and Fredenhagen, K.,
"Microlocal analysis and interacting quantum field 
theories: Renormalization on physical backgrounds", {\it Commun. Math. 
Phys.} {\bf 208} (2000) 623

\bibitem{BFV}Brunetti, R., Fredenhagen, K., and Verch, R.,''The 
generally covariant locality principle -- A new paradigm for local 
quantum physics'', {\it Commun. Math. Phys.} {\bf 237} 31-68

\bibitem{BOR}Buchholz, D., Ojima, I., Roos, H.,''Thermodynamic 
Properties of Non-Equilibrium States in Quantum Field Theory'',
{\it Annals Phys.} {\bf 297} (2002) 219-242

\bibitem{DF1}D\"utsch, M., and Fredenhagen, K.,''Algebraic Quantum Field 
Theory, Perturbation Theory, and the Loop Expansion'', 
{\it Commun. Math. Phys.}  {\bf 219} (2001) 5

\bibitem{DF2}D\"utsch, M. and Fredenhagen, K.,
''The Master Ward Identity and Generalized Schwinger-Dyson Equation
in Classical Field Theory'', {\it Commun. Math. Phys.}  
{\bf 243} (2003) 275-314

\bibitem{DF3}D\"utsch, M. and Fredenhagen, K., ``Causal perturbation theory 
in terms of retarded products, and a proof of the Action Ward Identity'',
to appear in {\it Rev. Math. Phys}

\bibitem{BD}D\"utsch, M. and Boas, F.-M.,
''The Master Ward Identity'', {\it Rev. Math. Phys} {\bf 14}
 (2002) 977-1049

\bibitem{EG}Epstein, H. and Glaser, V., "The role of locality in 
perturbation theory", {\it Ann. Inst. H. Poincar\'e A} {\bf 19} (1973) 211

\bibitem{GLZ} Glaser, V., Lehmann, H. and Zimmermann, W., ``Field
Operators and Retarded Functions'' \textit{Nuovo Cimen.} 
\textbf{6} (1957) 1122.

\bibitem{Haag} Haag, R., ``Local Quantum Physics: Fields, particles
  and algebras'', Springer-Verlag, Berlin, 2nd ed. (1996)

\bibitem{HW1-2} Hollands, S., Wald, R. M., ``Local Wick Polynomials
and Time-Ordered-Products of Quantum Fields in Curved Spacetime'',
{\it Commun. Math. Phys.}  {\bf 223} (2001) 289

Hollands, S., Wald, R. M., ``Existence of Local Covariant
Time-Ordered-Products of Quantum Fields in Curved Spacetime'',
{\it Commun. Math. Phys.} {\bf 231} (2002) 309-345

\bibitem{HW} Hollands, S., Wald, R. M., ``On the Renormalization Group
  in Curved Spacetime'', {\it Commun. Math. Phys.} {\bf 237} (2003) 123-160

\bibitem{HW4} Hollands, S., Wald, R. M., ``Conservation of the stress tensor in 
interacting quantum field theory in curved spacetimes'', gr-qc/0404074

\bibitem{Mar} Marecki, P., ``Quantum Electrodynamics on background external fields'',
hep-th/0312304

\bibitem{Pinter} Pinter, G.,''Finite Renormalizations in the
  Epstein-Glaser Framework and Renormalization of the $S$-Matrix
of $\Phi^4$-Theory'', {\it Ann. Phys. (Leipzig)} {\bf 10} (2001) 333

\bibitem{Scharf} Scharf, G., 
{\it "Finite Quantum Electrodynamics. The causal approach"}, 
2nd. ed., Springer-Verlag (1995)

\bibitem{S-wiley} Scharf, G., {\it ``Quantum Gauge Theories - 
A True Ghost Story''}, John Wiley and Sons (2001)

\bibitem{Ste} Steinmann, O., ``Perturbation expansions in axiomatic field 
theory'', Lecture Notes in Physics {\bf 11}, Berlin-Heidelberg-New York:
Springer-Verlag (1971)

\bibitem{SP} Popineau, G., and Stora, R.,
"A pedagogical remark on the main theorem of perturbative 
renormalization theory", unpublished preprint (1982)

\bibitem{AWI}Stora, R., ``Pedagogical Experiments in Renormalized
  Perturbation Theory'', contribution to the conference 'Theory of 
Renormalization and Regularization', Hesselberg, Germany (2002),
http://wwwthep.physik.uni-mainz.de/$\sim$scheck/Hessbg02.html;
and private communication

\bibitem{StPe}Stueckelberg, E.~C.~G., Petermann, A., ``La normalisation
  des constantes dans la theorie des quanta'', {\it Helv. Phys. Acta}
{\bf 26} (1953) 499-520


\end{thebibliography}
\end{document}